# Substrate Coupling Suppresses Size Dependence of Thermal Conductivity in Supported Graphene

Jie Chen,[1, *] Gang Zhang,[2, †] and Baowen Li[1, 3, ‡]

[1] Department of Physics, Centre for Computational Science and Engineering, and Graphene Research Centre, National University of Singapore, Singapore 117542

[2] Key Laboratory for the Physics and Chemistry of Nanodevices and Department of Electronics, Peking University, Beijing 100871, People's Republic of China

[3] NUS-Tongji Center for Phononics and Thermal Energy Science and School of Physical Science and Engineering, Tongji University, Shanghai 200092, People's Republic of China

Abstract: Thermal conductivity $\kappa$ of both suspended and supported graphene has been studied by using molecular dynamics simulations. Obvious length dependence is observed in $\kappa$ of suspended single-layer graphene (SLG), while $\kappa$ of supported SLG is insensitive to the length. The simulation result of room temperature $\kappa$ of supported SLG is in good agreement with experimental value. In contrast to the decrease in $\kappa$ induced by inter-layer interaction in suspended few-layer graphene (FLG), $\kappa$ of supported FLG is found to increase rapidly with the layer thickness, reaching about 90% of that of bulk graphite at six layers, and eventually saturates at the thickness of 13.4 nm. More interestingly, unlike the remarkable substrate dependent $\kappa$ in SLG, the effect of substrate on thermal transport is much weaker in FLG. The underlying physics is investigated and presented.

[*] Email: phychenj@nus.edu.sg
[†] Email: zhanggang@pku.edu.cn
[‡] Email: phylibw@nus.edu.sg

Size dependence of thermal conductivity has been reported in various low dimensional nanomaterials [1-7] as well as in many low dimensional lattice models [8, 9]. The most striking and controversial finding is that thermal conductivity in one dimensional systems (like nanotube and nanowire) diverges with length as power law, whereas it diverges logarithmically in two dimensional systems.

As a novel two dimensional material, graphene has been the focus of intense studies in recent years [10-12]. It is found that thermal conductivity of suspended graphene [5, 6] and graphene nanoribbons (GNRs) [7] is also size dependent. However, in many device applications, graphene sheets are usually supported by and integrated with substrates. It is therefore of primary interest to understand the size dependence of thermal conductivity in supported graphene.

In this paper, we systematically investigate the impacts of sample size, the thickness of graphene layers, and the substrate coupling strength on thermal conductivity of supported graphene on amorphous silicon dioxide ($SiO_2$) substrate. We employ molecular dynamics (MD) simulations in combination with phonon spectral analysis to elucidate the underlying physical mechanisms responsible for the heat conduction in supported graphene. Our study provides useful insights to better understand the thermal transport in supported graphene, which will be helpful in the design of graphene based devices.

In our study, all MD simulations are performed by using LAMMPS package [13]. Tersoff potential with optimized parameters for graphene [14] is adopted to model the intra-layer C-C interactions within the same graphene sheet. Tersoff parameter set for $SiO_2$ developed by Munetoh *et al.* [15] is used to model the interatomic interactions in the substrate. The inter-layer C-C interactions between adjacent graphene layers and the graphene-substrate interactions are model as van der Waal (vdW) type with

Lennard-Jones (LJ) potential $V(r_{ij}) = 4\chi\varepsilon_{ij}\left[\left(\frac{\sigma_{ij}}{r_{ij}}\right)^{12} - \left(\frac{\sigma_{ij}}{r_{ij}}\right)^{6}\right]$, where $r_{ij}$ is the interatomic distance, $\varepsilon_{ij}$ and $\sigma_{ij}$ are the bond-order force field parameters, and $\chi$ is a dimensionless scaling factor ($\chi=1$ by default). The LJ potential parameters for C-C bond are taken from Ref. [16], and the LJ parameters for C-Si and C-O bonds are taken from Ref. [17]. The cut-off distance in LJ potential is set as $2.5\sigma_{ij}$ for all kinds of bonds. Moreover, the neighbor list is dynamically updated every ten time steps, and each time step is set as 0.5 fs in our simulations.

The amorphous $SiO_2$ substrate is constructed by the standard high-temperature annealing process (see supplementary information for details), which has been widely used to simulate amorphous $SiO_2$ in literatures [18, 19]. Non-equilibrium MD simulation with Langevin heat bath [20] is used to study thermal transport in both suspended and supported graphene, as shown in Fig. 1. Periodic boundary condition is used in the width direction, and fixed boundary condition is used in the length direction. Thermal conductivity is calculated as $\kappa = -J/\nabla T$, where $\nabla T$ and $J$ is, respectively, the temperature gradient and the heat flux transported in graphene region only. For supported graphene, both the graphene and substrate are attached to the heat bath with the same temperature at each end (Fig. 1(b)), while only heat flux transported in graphene region is recorded in our simulations. More details about non-equilibrium MD simulation can be found in supplementary information. All the results reported in our study are ensemble averaged over six independent runs with different initial conditions.

To examine the size effect in MD simulations, we first calculate thermal conductivity of suspended single-layer graphene (SLG). In contrast to the obvious chirality effect on thermal conductivity of GNRs [7, 21], we find in our simulation negligible chirality effect on thermal conductivity of suspended SLG when the width is large enough, suggesting that the edge scatterings in the width direction are greatly suppressed by the periodic boundary condition. For instance, with fixed graphene size of $L$=300 Å and $W$=52 Å, the calculation results for suspended SLG are $\kappa_{zigzag}$=1009±42 W/m-K and $\kappa_{armchair}$=1002±41 W/m-K. Therefore, we fix the width as $W$=52 Å and only consider zigzag graphene in our simulations for the rest of our study.

Figure 1(c) shows thermal conductivity of suspended SLG versus the length at room temperature. We find thermal conductivity of suspended SLG increases with the length, and does not converge to a finite value even for the longest length considered in our study ($L$=200 nm). Similar phenomenon has also been observed by Guo *et al.* in GNRs with length less than 60 nm [7]. This length effect on thermal conductivity of suspended SLG found in our simulation is due to the intrinsically long phonon mean free path of graphene, which has the phonon mean free path up to 775 nm at room temperature [22] and thus leads to the high thermal conductivity of graphene.

When the SLG is put on amorphous $SiO_2$ substrate, the heat conduction properties have changed remarkably. Firstly, there exists negligible length effect on thermal conductivity of supported SLG compared to that of the suspended case (Fig. 1(c)). In the presence of the thick block of amorphous $SiO_2$ substrate, the computational cost is much higher than the suspended case. Therefore, the supported SLG considered in our study has a limited length up to 600 Å. In addition, thermal conductivity of supported SLG is much lower than that of suspended SLG with the same length. For instance, with the length $L$=300 Å, thermal conductivity of suspended SLG is 1009±42 W/m-K, while that of $SiO_2$-supported SLG is only 609±19 W/m-K, showing a 40% reduction of thermal conductivity.

It should be noted that our calculation result for supported SLG is in a very good agreement with a recent experimental result around 600 W/m-K at room temperature [23]. Table I shows that our calculation results for supported SLG have very weak dependence on the thickness $D$ of $SiO_2$ substrate, or the separation distance $d$ between graphene and substrate, which is defined in our simulation as the minimum distance between graphene and substrate atoms along the out-of-plane ($z$) direction (Fig. 1(b)).

To understand this distinct size dependence of thermal conductivity in suspended and supported SLG, we carry out the spectral energy density (SED) analysis $\Phi(k, \omega)$ [24] for phonons in graphene (see supplementary information for calculation details), where $k$ is the wavevector index, $\omega$ is the angular frequency.

The SED is a very useful tool to extract phonon information (dispersion relation and lifetime) from MD simulation which can incorporate the full anharmonicity of the atomic interactions, and has been used in a number of studies [19, 24, 25].

Previous theoretical study has suggested that thermal conductivity of graphene is dominated by contributions from ZA phonons [26], which have an anomalously large density of states near zone-center compared to the in-plane modes. Therefore, we focus on the zone-center ZA phonons in the SED analysis.

Figure 2a shows the normalized SED for the zone-center ($k$=1) ZA modes in both suspended SLG and supported graphene with different layers. The location of the peak denotes the phonon eigen-frequency, and the width of the peak is inversely proportional to the phonon lifetime [24]. For suspended SLG, only one distinct and narrow peak shows up, which corresponds to ZA phonon with the lowest frequency allowed in the system. After putting on the substrate, the ZA phonon shifts to high frequency and the zone-center peak is notably broaden compared to the suspended case (inset of Fig. 2a), with more than two orders of magnitude reduction in SED intensity (Fig. 2b). This suggests that ZA phonons are greatly suppressed by the graphene-substrate interaction in supported SLG. The blue-shift of ZA phonon spectrum indicates that the graphene-substrate interaction can induce additional (apart from boundary scattering) low-frequency cut-off, and thus hinders the heat transport in supported graphene. Similar lift-up of phonon spectrum has also been observed in one dimensional lattice model with on-site potential [27].

The effects of broadening and lift-up of phonon spectrum induced by substrate coupling greatly suppress the thermal transport of long wavelength phonons, leading to the significantly reduced thermal conductivity and negligible size effect on thermal transport in supported SLG.

Furthermore, we notice that additional phonon modes which are otherwise absent in suspended SLG are introduced into the supported graphene (Fig. 2(a)) due to the graphene-substrate interaction. These modes have nearly zero group velocity due to the amorphous nature of $SiO_2$ [28] and thus contribute little to thermal transport.

So far we have explored the length dependence of thermal conductivity in both suspended and supported SLG. In the next part, we extend our discussion to the thickness effect of thermal conductivity in supported few-layer graphene (FLG). In the following simulations, we fix the substrate thickness $D$=20 Å and separation distance $d$=2 Å, the in-plane size of graphene at width $W$=52 Å and length $L$=300 Å, and tune the thickness of FLG.

Figure 3 shows the room temperature thermal conductivity of supported FLG versus the number of layers. The maximum thickness considered in our study is 12 layers, which has a total number of 91,496 atoms including the substrate atoms. With the increase of graphene layers, it exhibits a very rapid enhancement in thermal conductivity of supported FLG at the beginning, followed by a much slower increase after six layers, showing an obvious two-stage increase characteristic which has also been observed in a recent experimental study on the thickness dependent thermal conductivity of encased graphene [29].

On the other hand, it has been reported that thermal conductivity of suspended FLG decreases with the thickness due to the inter-layer interactions [30, 31]. It is known that both inter-layer interaction and graphene-substrate interaction can lead to the reduction of thermal conductivity in graphene due to damping of the ZA phonons. However, the impacts from these two types of interactions are different. In principle, both of them can be described by vdW form potentials. From the force field parameter of LJ potential used in our simulations, we find $\varepsilon_{C-Si}/\varepsilon_{C-C}=3.72$ and $\varepsilon_{C-O}/\varepsilon_{C-C}=1.44$, suggesting a stronger interaction strength between graphene and $SiO_2$ substrate compared to the graphene inter-layer interaction. Thus the graphene-substrate interaction has more remarkable impact on the reduction of thermal conductivity than the inter-layer interaction does. In the supported FLG, because the upper layers have less interaction (close to zero) with the substrate, the impact of substrate on thermal transport in FLG becomes weaker with the increase of the layer thickness.

This point can be clearly seen from SED shown in Fig. 2. With the increase of layers in supported graphene, the ZA peak shifts back to low frequency, and the width of the ZA peak is reduced compared to that of supported SLG. Meanwhile, the intensity of the additional phonon modes arising from the graphene-substrate coupling is suppressed.

All these findings suggest that when the number of layers increases in supported FLG, the impact of substrate coupling reduces correspondingly. This conclusion is further supported by Fig. 2b, which shows the greatly reduced SED intensity in supported SLG increases with the number of layers.

As the effects of broadening and lift-up of phonon spectrum induced by substrate coupling are weaken with the increase of graphene thickness, thermal conductivity of supported FLG increases correspondingly. Our simulation results reveal that although thermal conductivity of graphene is reduced remarkably by substrate coupling, one can enhance its thermal conductivity by increasing the thickness. Thus multi-layer graphene is more favorable than its single-layer counterpart for heat dissipation in practical device application on substrate.

We further estimate the thermal conductivity in the bulk graphite limit based on the double exponential fitting to our existing data (see supplementary information for details). The double exponential fitting can not only qualitatively represent the two-stage increase characteristic shown in Fig. 3, but also provide a self-consistent estimation of thermal conductivity in the bulk graphite limit ($n\rightarrow\infty$). The estimated thermal conductivity in bulk graphite limit is $\kappa_{graphite}$=962 W/m-K. The recorded room temperature thermal conductivity of bulk graphite in literature ranges from 800 W/m-K for regular bulk graphite to 2000 W/m-K for high-quality bulk graphite [30]. Our estimated value is within this range, although below the high-quality bulk graphite limit presumably owing to the onset of the phonon-boundary scattering in our simulations.

Using the estimated $\kappa_{graphite}$ as reference, we show self-consistently in the inset of Fig. 3 the normalized thermal conductivity of supported FLG versus the number of layers based on the double exponential fit formula. Due to the rapid increase of thermal conductivity from SLG, thermal conductivity of the supported six-layer graphene already reaches 90% of that of the bulk graphene limit.

With the further increase of thickness, thermal conductivity of supported FLG is found to saturate to that of the bulk graphite limit at the thickness of 40 layers (13.4 nm). This saturation thickness of 13.4 nm predicted in our study is about half of that in the encased graphene (25 nm) estimated from a model based on experimental results in Ref. [29]. This is because in the encased graphene, both top and bottom substrates have interactions with the FLG in the middle, $\kappa$ is reduced compared to our current model with the same thickness [29], thus leading to a larger saturation thickness in the encased graphene.

Taking amorphous $SiO_2$ supported graphene as an example, we have studied the impacts of graphene-substrate and layer-layer interaction on graphene thermal conductivity. In practical application, other substrates, such as boron nitride, are also commonly used. Thus it is highly desirable to explore the dependence of heat conduction properties on graphene-substrate interaction strength, which can facilitate the choice of substrate for high-performance graphene based planar heat dissipation. To this end, we tune the strength of graphene-substrate interaction in the LJ potential, and redo our simulation to calculate the corresponding thermal conductivity of supported graphene.

Figure 4 shows the effect of graphene-substrate coupling strength on thermal conductivity of single-layer and five-layer supported graphene. Stronger substrate bonding is found to result in further reduction of thermal conductivity in supported graphene. This is due to the increase of graphene-substrate coupling strength can significantly shorten phonon lifetime and thus lead to reduction of thermal conductivity.

Our simulation results are in line with a recent theoretical analysis on the phonon lifetime in suspended and supported graphene [19]. Furthermore, we find the impact of graphene-substrate interaction strength is different for SLG and FLG. For instance, when $\chi$ increases from 1 to 2, thermal conductivity of SLG reduces from 609±19

W/m-K to 482±16 W/m-K, showing about 20% reduction. For five-layer graphene, thermal conductivity changes from 849±24 W/m-K to 780±18 W/m-K, with only about 10% reduction. Thus for the high-performance graphene based heat dissipation application, the choice of substrate which corresponds to the different coupling strength is very critical for SLG. However, the impact of substrate is much weaker for FLG.

In conclusion, by using molecular dynamics simulations, we have studied thermal conductivity of suspended and supported graphene at room temperature. We find thermal conductivity of suspended SLG increases with the length, and does not converge to a finite value within the length of 200 nm. However, thermal conductivity of $SiO_2$ supported SLG is much lower than that of its suspended counterpart, and is almost independent on the length. In contrast to the decrease of thermal conductivity induced by inter-layer interaction in suspended FLG, thermal conductivity of supported FLG is found to increase rapidly with the layer thickness, reaching about 90% of that of bulk graphite limit at 6 layers, and eventually saturates at the thickness of 13.4 nm (40 layers). Moreover, we find that compare to SLG, the effect of substrate coupling on thermal transport is much weaker in FLG.

The phonon spectral analysis reveals that the graphene-substrate interaction causes remarkable damping of the flexural phonons, leading to the significantly reduced thermal conductivity in SLG after put on substrate. With the increase of layer thickness, this substrate-induced damping becomes weaker and thus leads to the increase of thermal conductivity in supported FLG. Our results are helpful to identify the roles of graphene-substrate and inter-layer interaction on thermal transport in supported few-layer graphenes, and thereby facilitate the development of practical high-performance graphene based devices for nanoscale heat dissipation and thermal management.

## Acknowledgements

The work has been supported by a grant W-144-000-305-112 MOE, Singapore. G.Z. was supported by the Ministry of Education of China (Grant No. 20110001120133) and the Ministry of Science and Technology of China (Grant Nos. 2011CB933001). J.C. acknowledges the World Future Foundation (WFF) for awarding him the WFF PhD Prize in Environmental and Sustainability Research (2012) and the financial support to this work. J.C. acknowledges the helpful discussion with Alan J. H. McGaughey about the spectral energy density analysis.

**Reference**

[1] S. Maruyama, *Physica B*, 2002, **323**, 193.

[2] G. Zhang and B. Li, *J. Chem. Phys.*, 2005, **123**, 114714.

[3] C.W. Chang, D. Okawa, H. Garcia, A. Majumdar and A. Zettl, *Phys. Rev. Lett.*, 2008, **101**, 075903.

[4] N. Yang, G. Zhang and B. Li, *Nano Today*, 2010, **5**, 85.

[5] Z. Wang, R. Xie, C.T. Bui, D. Liu, X. Ni, B. Li and J. T. L. Thong, *Nano Lett.*, 2011, **11**, 113.

[6] D. L. Nika, A. S. Askerov and A. A. Balandin, *Nano Lett.*, 2012, **12**, 3238.

[7] Z. Guo, D. Zhang and X.-G. Gong, *Appl. Phys. Lett.*, 2009, **95**, 163103.

[8] A. Dhar, *Adv. Phys.*, 2008, **57**, 457.

[9] S. Liu, X. Xu, R. Xie, G. Zhang and B. Li, *Eur. Phys. J. B*, 2012, **85**, 337.

[10] K. S. Novoselov, A. K. Geim, S. V. Morozov, D. Jiang, Y. Zhang, S.V. Dubonos, I. V. Grigorieva and A. A. Firsov, *Science*, 2004, **306**, 666.

[11] Y. Zhang, Y. -W. Tan, H. L. Stormer and P. Kim, *Nature*, 2005, **438**, 201.

[12] A. A. Balandin, *Nat. Mater.*, 2011, **10**, 569.

[13] S. Plimpton, *J. Comput. Phys.*, 1995, **117**, 1.

[14] L. Lindsay and D. A. Broido, *Phy. Rev. B*, 2010, **81**, 205441.

[15] S. Munetoh, T. Motooka, K. Moriguchi and A. Shintani, *Comput. Mater. Sci.*, 2007, **39**, 334.

[16] L. A. Girifalco, M. Hodak and R. S. Lee, *Phys. Rev. B*, 2000, **62**, 13104.

[17] A. K. Rappe, C. J. Casewit, K. S. Colwell, W. A. Goddard and W. M. Skiff, *J. Am. Chem. Soc.*, 1992, **114**, 10024.

[18] J. Chen, G. Zhang and B. Li, *J. Appl. Phys.*, 2012, **112**, 064319.

[19] B. Qiu and X. Ruan, *Appl. Phys. Lett.*, 2012, **100**, 193190.

[20] J. Chen, G. Zhang and B. Li, *J. Phys. Soc. Jpn.*, 2010, **79**, 074604.

[21] J. Hu, X. Ruan and Y. P. Chen, *Nano Lett.*, 2009, **9**, 2730.

[22] S. Ghosh, I. Calizo, D. Teweldebrhan, E. P. Pokatilov, D. L. Nika, A. A. Balandin, W. Bao, F. Miao and C. N. Lau, *Appl. Phys. Lett.*, 2008, **92**, 151911.

[23] J. H. Seol, I. Jo, A. L. Moore, L. Lindsay, Z. H. Aitken, M. T. Pettes, X. Li, Z. Yao, R. Huang, D. Broido, N. Mingo, R. S. Ruoff and Li Shi, *Science* 2010, **328**, 213.

[24] J. A. Thomas, J. E. Turney, R. M. Iutzi, C. H. Amon and A. J. H. McGaughey, *Phys. Rev. B*, 2010, **81**, 081411(R).

[25] Z. -Y. Ong and E. Pop, *Phys. Rev. B*, 2011, **84**, 075471.

[26] L. Lindsay, D. A. Broido and N. Mingo, *Phys. Rev. B*, 2010, **82**, 115427.

[27] B. Li, L. Wang and G. Casati, *Phys. Rev. Lett.*, 2004, **93**, 184301.

[28] D. Donadio and G. Galli, *Phys. Rev. Lett.*, 2009, **102**, 195901.

[29] W. Jiang, Z. Chen, W. Bao, C. N. Lau and C. Dames, *Nano Lett.*, 2010, **10**, 3909.

[30] S. Ghosh, W. Bao, D. L. Nika, S. Subrina, E. P. Pokatilov, C. N. Lau and A. A. Balandin, *Nat. Mater.*, 2010, **9**, 555.

[31] W.-R. Zhong, M.-P. Zhang, B.-Q. Ai and D.-Q. Zheng, *Appl. Phys. Lett.*, 2011, **98**, 113107.

Table I. Room temperature thermal conductivity of supported single-layer graphene with different sizes of simulation domain. Here zigzag graphene with fixed width $W$=52 Å and length $L$=300 Å is used in the calculations.

| Simulation domain | $\kappa$ (W/m-K) |
|---|---|
| d=2 Å, D=20 Å | 609±19 |
| d=2 Å, D=30 Å | 598±21 |
| d=3.35 Å, D=20 Å | 616±21 |

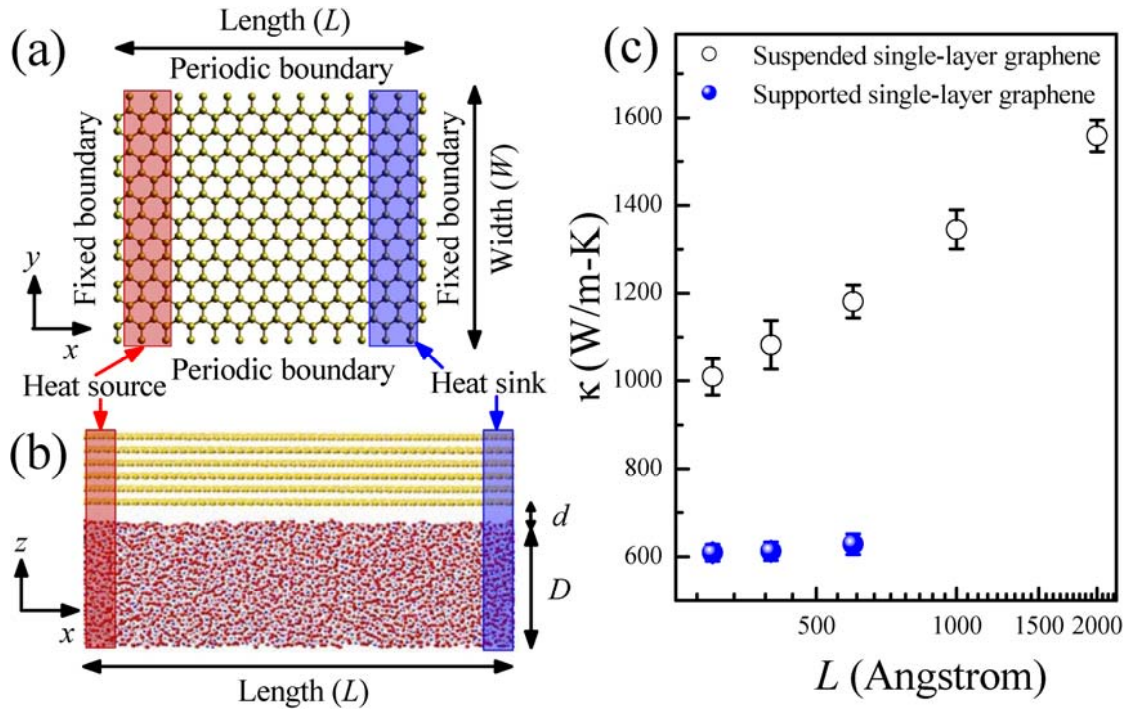

Figure 1. Schematic picture of simulation setup and thermal conductivity of single-layer graphene at room temperature. (a) Top-view of suspended single-layer graphene. (b) Side-view of supported few-layer graphene on $SiO_2$ substrate. The parameter $D$ and $d$ denote the thickness of the substrate and the separation distance between graphene and substrate, respectively. (c) Thermal conductivity $\kappa$ of suspended (empty circle) and supported (solid circle) single-layer graphene versus the length $L$ at room temperature. Here zigzag graphene with fixed width $W$=52 Å is used for both suspended and supported cases. For the supported graphene, the thickness of $SiO_2$ substrate is $D$=20 Å with a separation distance $d$=2 Å.

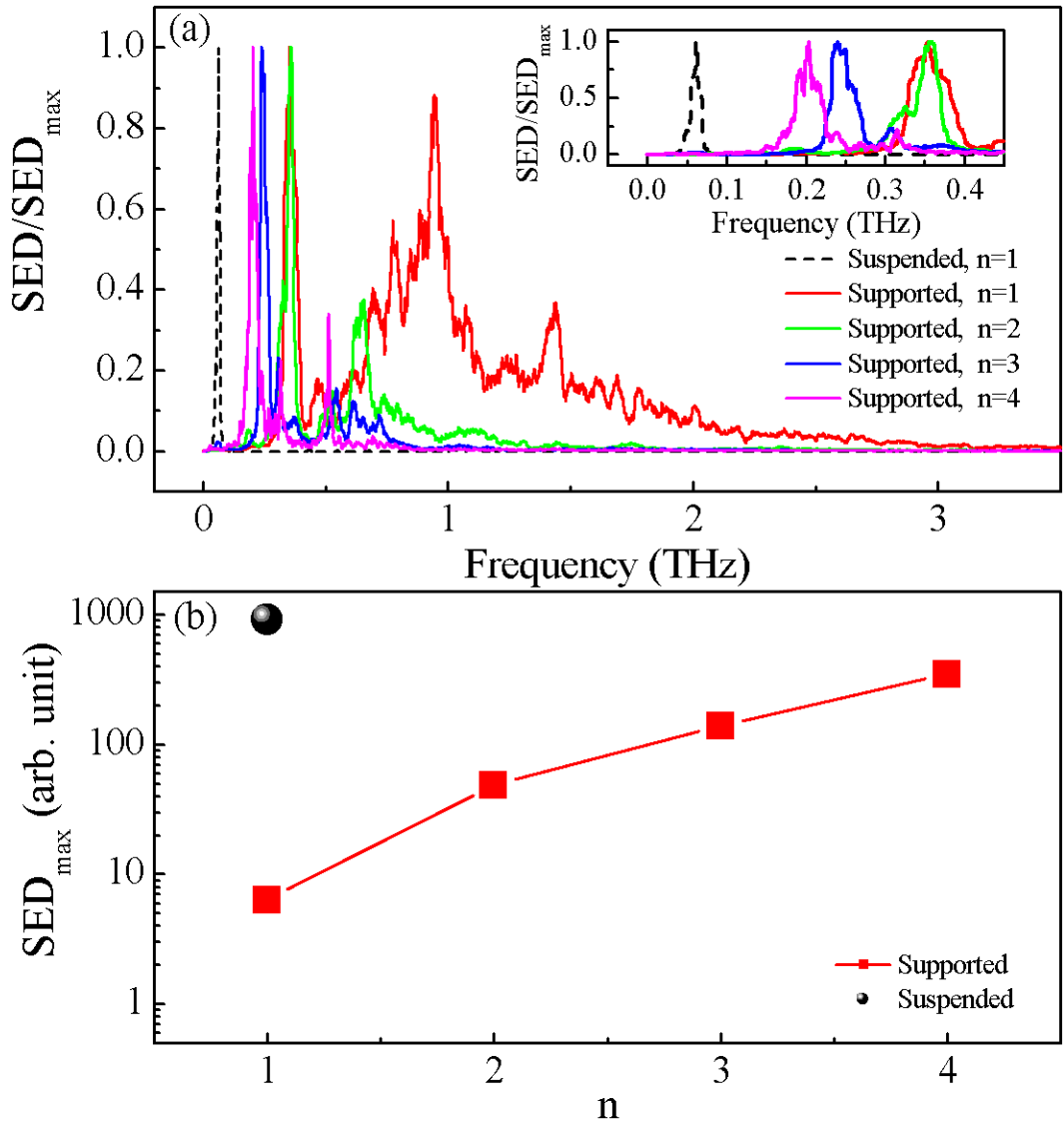

Figure 2. Spectral energy density (SED) analysis for zone-center ZA phonons ($k$=1) in both suspended and supported graphene. (a) Normalized SED for suspend SLG (dashed line) and supported graphene with different layers (solid lines). The inset zooms in for the low frequency ZA peaks near zone-center. (b) SED intensity of the low frequency ZA peaks for suspend SLG (circle) and supported graphene with different layers (square).

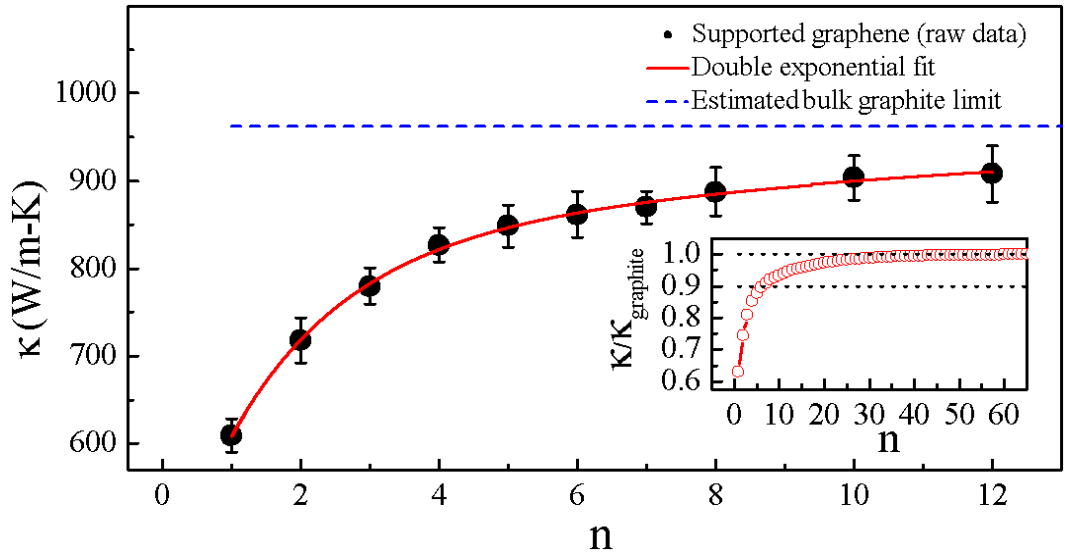

Figure 3. Room temperature thermal conductivity $\kappa$ of supported few-layer graphene versus the number of layers $n$. Here zigzag graphene with fixed width $W$=52 Å and length $L$=300 Å is used in the calculations. The solid circle draws the raw data from MD simulation, and the solid line draws the double exponential fit. The dashed line draws the estimated bulk graphite limit for reference, and the inset shows the normalized $\kappa$. The dotted lines are drawn to guide the eye.

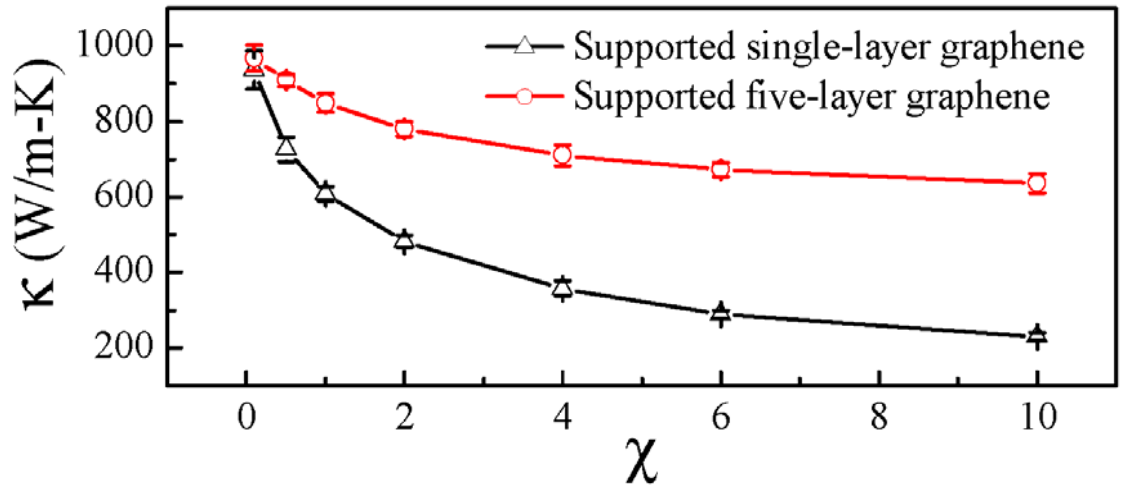

Figure 4. Effect of graphene-substrate coupling strength ($0.1 \leq \chi \leq 10$) on room temperature thermal conductivity of single-layer (triangle) and five-layer (circle) supported graphene. Here zigzag graphene with fixed width $W$=52 Å and length $L$=300 Å is used in the calculations.

# Supplementary information for "Substrate Coupling Suppresses Size Dependence of Thermal Conductivity in Supported Graphene"

Jie Chen,[1, *] Gang Zhang,[2, †] and Baowen Li[1, 3, ‡]

[1] Department of Physics, Centre for Computational Science and Engineering, and Graphene Research Centre, National University of Singapore, Singapore 117542

[2] Key Laboratory for the Physics and Chemistry of Nanodevices and Department of Electronics, Peking University, Beijing 100871, People's Republic of China

[3] NUS-Tongji Center for Phononics and Thermal Energy Science and School of Physical Science and Engineering, Tongji University, Shanghai 200092, People's Republic of China

## I. HIGH-TEMPERATURE ANNEALING PROCESS

To construct the amorphous $SiO_2$ substrate at room temperature, we start with the crystalline form of $SiO_2$: alpha-quartz ($\alpha$-quartz). We apply Langevin heat bath to equilibrate $\alpha$-quartz at 3000 K (above melting point) for 100 ps in order to achieve the amorphous structure. The resultant structure is then annealed to room temperature with a constant cooling rate of $10^{13}$ K/s [1]. The partial pair distribution function [2] for different chemical bonds in the amorphous $SiO_2$ structure generated in our study (Fig. S1) shows excellent agreement with previous study [3], which highlights the accuracy of the high-temperature annealing process used in our simulation.

[*] Email: phychenj@nus.edu.sg
[†] Email: zhanggang@pku.edu.cn
[‡] Email: phylibw@nus.edu.sg

## II. NON-EQUILIBRIUM MOLECULAR DYNAMICS SIMULATION

Before non-equilibrium MD simulation, the canonical ensemble MD simulation with Langevin heat bath first runs for $10^5$ steps to equilibrate the whole system at room temperature. After structure relaxation, fixed boundary condition is used at the two ends of the length (*x*) direction (Fig. 1a). Next to the fixed boundary, Langevin heat baths with different temperature are applied to the two ends of *x* direction to simulate the heat source (red box) and heat sink (blue box) in real experiment, respectively. Periodic boundary condition is used in the width (*y*) direction, and free boundary condition is used in the out-of-plane (*z*) direction. Both the graphene and substrate are attached to the heat bath with the same temperature at each end (Fig. 1(b)). The non-equilibrium MD simulations are then performed long enough ($10^7$ time steps) to allow the system to reach the non-equilibrium steady state where the temperature gradient is well established and the heat flux passing through the system is time independent. Thermal conductivity is calculated according to

$$\kappa = -J/\nabla T, \quad \quad (S1)$$

where $\nabla T$ and $J$ is, respectively, the temperature gradient and the heat flux transported in graphene region only. In our simulation, the heat flux is calculated according to the energy injected into / extracted from the heat source / heat sink in the graphene sheets only (exclude the heat flux in substrate) across unit area per unit time. These two rates are equal in the non-equilibrium steady state. For supported graphene, only heat flux transported in graphene region is recorded in our simulations. The cross section area (*S*) of the graphene sheets is defined as $S=3.35*W*n$ Å$^2$ in our calculations, where *W* is the width of graphene, and *n* is the number of layers. The temperature gradient is calculated according to the slope of the linear fit line of the local temperature in graphene along *x* direction.

## III. COMPARE THERMAL CONDUCTIVITY OF SUSPENDED SINGLE-LAYER GRAPHENE WITH LITERATURE VALUE

We notice that $\kappa$ of suspended SLG found in our study (~1000 W/m-K) is much larger than that reported by Ong *et al.* [4] (256 W/m-K) under the similar sample size (length is about 30 nm and width is about 5 nm). We contribute this discrepancy to the choice of interatomic potential used for graphene. The original Brenner potential was used in Ref. [4], while we use the Tersoff potential with optimized parameters set for graphene developed by Lindsay *et al.* [5], which can describe more accurately the upper optic phonon branches while providing a good fit to the acoustic velocity and phonon frequency. Lindsay *et al.* found that the original Brenner potential failed to accurately represent the zone-center velocities for all the acoustic modes [5], leading to 30% underestimation for longitudinal acoustic (LA) branch and 12% underestimation for transverse acoustic (TA) branch compared to experimental values. Since thermal conductivity depends critically on the group velocity of acoustic phonons based on the Boltzmann transport equation approach, Lindsay *et al.* found the use of original Brenner potential can lead to a large underestimation of thermal conductivity for suspended SLG [5].

## IV. SPECTRAL ENERGY DENSITY ANALYSIS

The spectral energy density (SED) in our calculation is defined as [4, 6]

$$\Phi(k,\omega) = \frac{1}{4\pi\tau_0 N_x N_y N_z} \sum_\alpha \sum_{b=1}^{B} m_b \left| \int_0^{\tau_0} \sum_{n_x=0}^{N_x-1} \sum_{n_y=0}^{N_y-1} \sum_{n_z=0}^{N_z-1} v_{\alpha,b}(n_x, n_y, n_z, t) \exp\left[\frac{2\pi i k n_x}{N_x} - i\omega t\right] dt \right|^2,$$

(S2)

where $k$ is the wavevector index, $\omega$ is the angular frequency, $\tau_0$ is the total simulation time, $\alpha$ is the Cartesian index, $b$ is the atom index in each unit cell, $m$ and $v$ is the atomic mass and velocity, respectively, and $N_x$, $N_y$, $N_z$ denotes the number of unit cell in $x$, $y$, $z$ direction, respectively. Here we consider a one-dimensional Brillouin zone along the length ($x$) direction ($1 \leq k \leq N_x$). We choose the same four-atom unit cell described in Ref. [4], and use a fixed simulation domain of $N_x$=20 and $N_y$=6 for the in-plane direction. Periodic boundary condition is used in both $x$ and $y$ (in-plane) direction, and free boundary condition is used in $z$ (out-of-plane) direction. After structure relaxation with Langevin heat bath, we carry out microcanonical ensemble (NVE) MD simulation to the whole system for 3 ns and 1 ns for the suspended and supported graphene, respectively, and record the velocity for each atom in the graphene every 5 fs. We further extend the NVE MD simulation steps and find the calculation results of SED is well converged within the abovementioned total simulation time. To examine the accuracy of our simulation, we calculate the SED for LA phonons near zone-center ($k$=1) in suspended SLG and find the eigen-frequency at $f$=4.45±0.05 THz (Fig. S2). Using the lattice constant $L_0$ and group velocity for LA phonons $v_{LA}$ associated with the optimized Tersoff potential listed in Ref. [5], the theoretical estimation gives rise to $f=v_{LA}/(L_0 N_x)$=4.38 THz, in good agreement with our simulation results.

## V. ESTIMATE THERMAL CONDUCTIVITY OF BULK GRAPHITE

In order to estimate the thermal conductivity in the bulk graphite limit from our existing data, we fit the raw data from MD simulation according to the double exponential function based on the two-stage increase characteristic as

$$\kappa = \kappa_0 + A_1\left(1 - e^{-n/B_1}\right) + A_2\left(1 - e^{-n/B_2}\right), \tag{S3}$$

where $\kappa_0$, $A_1$, $B_1$, $A_2$ and $B_2$ are fitting parameters. The double exponential fitting yields $\kappa_0$=417, $A_1$=147, and $A_2$=398, suggesting thermal conductivity in bulk graphite limit as $\kappa_{graphite} = \kappa_0 + A_1 + A_2$=962 W/m-K.

**Reference**

[1] S. Munetoh, T. Motooka, K. Moriguchi and A. Shintani, *Comput. Mater. Sci.*, 2007, **39**, 334.

[2] K. Matsunaga and Y. Iwamoto, *J. Am. Ceram. Soc.*, 2001, **84**, 2213.

[3] W. Jin, R. K. Kalia, P. Vashishta and J. P. Rino, *Phys. Rev. Lett.*, 1993, **71**, 3146.

[4] Z. -Y. Ong and E. Pop, *Phys. Rev. B*, 2011, **84**, 075471.

[5] L. Lindsay and D. A. Broido, *Phy. Rev. B*, 2010, **81**, 205441.

[6] J. A. Thomas, J. E. Turney, R. M. Iutzi, C. H. Amon and A. J. H. McGaughey, *Phys. Rev. B*, 2010, **81**, 081411(R).

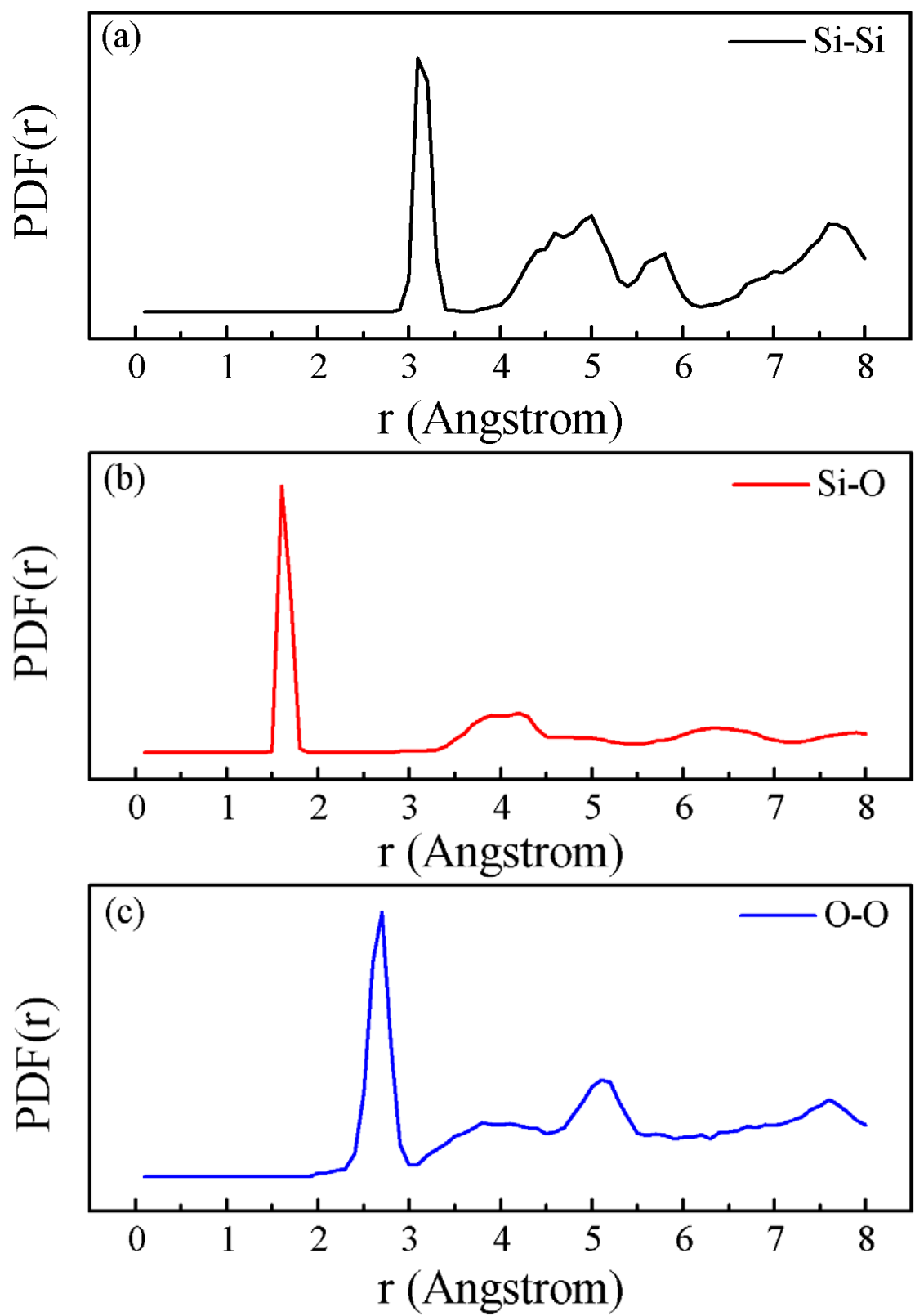

Figure S1. Partial pair distribution function (PDF) for different chemical bonds in amorphous $SiO_2$ at 300 K. (a) Si-Si bond. (b) Si-O bond. (c) O-O bond.

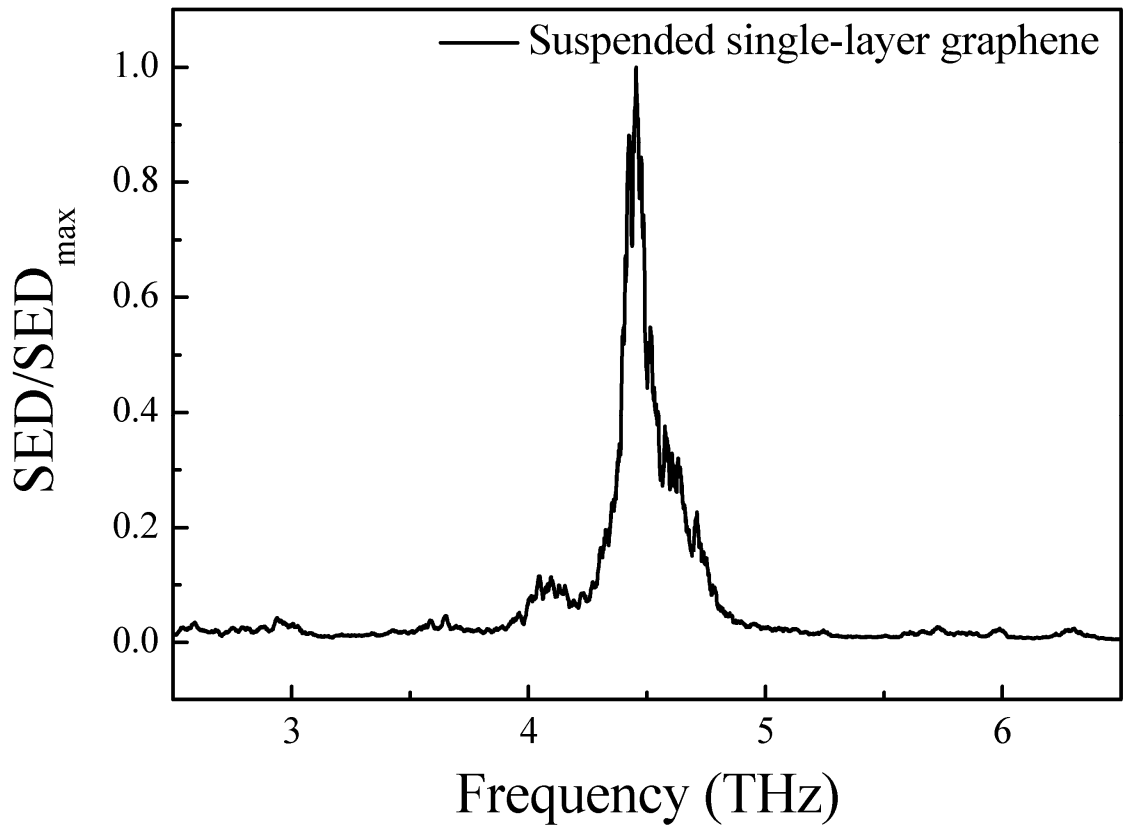

Figure S2. Normalized spectral energy density (SED) for zone-center LA phonons ($k$=1) in suspended single-layer graphene.